\documentclass{aastex701}

\usepackage{tabularx}
\usepackage{array}

\makeatletter
\renewcommand{\frontmatter@title@above}{}
\makeatother

\begin{document}

\title{A Proliferated Space Architecture for Time-Domain Astrophysics}

\author[orcid=0000-0001-9201-4706,sname='Kocevski']{Daniel Kocevski}
\affiliation{NASA Marshall Space Flight Center}
\email[hide]{daniel.kocevski@nasa.gov}  

\author[orcid=0000-0002-2942-3379,sname='Burns']{Eric Burns}
\affiliation{Department of Physics \& Astronomy, Louisiana State University}
\email[hide]{noemail@email.org}  

\author[orcid=0000-0001-7139-2724,sname='Barclay']{Thomas Barclay}
\affiliation{NASA Goddard Space Flight Center}
\email[hide]{noemail@email.org}  

\author[orcid=0000-0002-9408-3964,sname='White']{Nicholas E. White}
\affiliation{Department of Physics, George Washington University}
\email[hide]{noemail@email.org}  

\author[orcid=0000-0003-3702-7592,sname='Galeazzi']{Massimiliano Galeazzi}
\affiliation{Department of Physics, University of Miami}
\email[hide]{noemail@email.org}  

\author[orcid=0000-0002-9700-0036,sname='O''Connor']{Brendan O'Connor}
\affiliation{Department of Physics, Carnegie Mellon University}
\email[hide]{noemail@email.org}  

\author[orcid=0000-0002-7851-9756,sname='Lien']{Amy Lien}
\affiliation{Department of Physics and Astronomy, University of Tampa}
\email[hide]{noemail@email.org}  

\author[orcid=0000-0003-2624-0056,sname='Fryer']{Chris Fryer}
\affiliation{Los Alamos National Laboratory}
\email[hide]{noemail@email.org}  

\author[sname='Fausey']{Hallie Fausey}
\affiliation{Department of Physics and Astronomy, Baylor University}
\email[hide]{noemail@email.org}  

\author[orcid=0000-0001-6191-1244,sname='McBride']{Fe McBride}
\affiliation{Physics and Astronomy, Bowdoin College}
\email[hide]{noemail@email.org}  

\author[orcid=0000-0002-7435-0869,sname='Cornish']{Neil Cornish}
\affiliation{Department of Physics, Montana State University}
\email[hide]{noemail@email.org}  

\author[orcid=0000-0003-1386-7861,sname='Pasham']{Dheeraj Pasham}
\affiliation{Department of Physics, George Washington University}
\email[hide]{noemail@email.org}  

\author[orcid=0000-0001-6654-5378,sname='Kuntz']{Kip Kuntz}
\affiliation{Department of Physics and Astronomy, Johns Hopkins University}
\email[hide]{noemail@email.org}  

\author[orcid=0000-0002-2028-9329,sname='Nugent']{Anya Nugent}
\affiliation{Center for Astrophysics \textbar~Harvard \& Smithsonian}
\email[hide]{noemail@email.org}  

\author[sname='Porter']{Scott Porter}
\affiliation{NASA Goddard Space Flight Center}
\email[hide]{noemail@email.org}  

\author[orcid=0000-0001-5929-4187,sname='Chang']{Tzu-Ching Chang}
\affiliation{Jet Propulsion Laboratory, California Institute of Technology}
\email[hide]{noemail@email.org}  

\author[orcid=0000-0002-3950-9598,sname='Lidz']{Adam Lidz}
\affiliation{Department of Physics and Astronomy, University of Pennsylvania}
\email[hide]{noemail@email.org}  

\author[orcid=0000-0001-9149-6707,sname='van der Horst']{Alexander van der Horst}
\affiliation{Department of Physics, George Washington University}
\email[hide]{noemail@email.org}  

\author[orcid=0000-0001-5780-8770,sname='Guiriec']{Sylvain Guiriec}
\affiliation{Department of Physics, George Washington University}
\email[hide]{noemail@email.org}  

\author[orcid=0000-0002-8294-9281,sname='Cackett']{Ed Cackett}
\affiliation{Department of Physics and Astronomy, Wayne State University}
\email[hide]{noemail@email.org}  

\author[orcid=0000-0002-8028-0991,sname='Hartmann']{Dieter Hartmann}
\affiliation{Department of Physics and Astronomy, Clemson University}
\email[hide]{noemail@email.org}  

\author[orcid=0000-0003-1673-970X,sname='Cenko']{Brad Cenko}
\affiliation{NASA Goddard Space Flight Center}
\email[hide]{noemail@email.org}  

\author[orcid=0000-0002-0468-6025,sname='Hui']{C. Michelle Hui}
\affiliation{NASA Marshall Space Flight Center}
\email[hide]{noemail@email.org}  

\author[orcid=0000-0002-4299-2517,sname='Parsotan']{Tyler Parsotan}
\affiliation{NASA Goddard Space Flight Center}
\email[hide]{noemail@email.org}  

\author[orcid=0000-0002-8262-2924,sname='Coughlin']{Michael W. Coughlin}
\affiliation{School of Physics and Astronomy, University of Minnesota}
\email[hide]{noemail@email.org}  

\author[orcid=0009-0005-0762-4507,sname='Neights']{Eliza Neights}
\affiliation{Department of Physics, George Washington University}
\affiliation{NASA Goddard Space Flight Center}
\email[hide]{noemail@email.org}  

\author[orcid=0000-0002-9249-0515,sname='Wadiasingh']{Zorawar Wadiasingh}
\affiliation{Department of Astronomy, University of Maryland}
\affiliation{NASA Goddard Space Flight Center}
\email[hide]{noemail@email.org}  

\author[sname='Grant']{Catherine Grant}
\affiliation{MIT Kavli Institute for Astrophysics and Space Research, Massachusetts Institute of Technology}
\email[hide]{noemail@email.org}  

\author[orcid=0000-0002-4593-0792,sname='Mukherjee']{Oindabi Mukherjee}
\affiliation{Universities Space Research Association}
\email[hide]{noemail@email.org}  

\author[orcid=0000-0003-2501-2270,sname='Longo']{Francesco Longo}
\affiliation{Department of Physics, University of Trieste}
\affiliation{INFN Trieste}
\email[hide]{noemail@email.org}  

\author[orcid=0000-0001-8018-5348,sname='Bellm']{Eric Bellm}
\affiliation{Department of Astronomy, University of Washington}
\email[hide]{noemail@email.org}  

\author[orcid=0000-0002-1018-9383,sname='Habig']{Alec Habig}
\affiliation{Department of Physics and Astronomy, University of Minnesota Duluth}
\email[hide]{noemail@email.org}  

\author[orcid=0000-0002-7991-028X,sname='Younes']{George Younes}
\affiliation{Department of Physics, University of Maryland, Baltimore County}
\affiliation{NASA Goddard Space Flight Center}
\email[hide]{noemail@email.org}  

\author[orcid=0000-0001-5624-2613,sname='Page']{Kim Page}
\affiliation{School of Physics \& Astronomy, University of Leicester}
\email[hide]{noemail@email.org}  

\author[sname='Sharma']{Shivam Kumar Sharma}
\affiliation{STIG/PhysPAG-NASA}
\email[hide]{noemail@email.org}  

\author[sname='Roberts']{Oliver J. Roberts}
\affiliation{School of Natural Sciences, University of Galway, Ireland}
\email[hide]{noemail@email.org}  

\author[orcid=0000-0003-0761-6388,sname='Hamburg']{Rachel Hamburg}
\affiliation{Universities Space Research Association}
\email[hide]{noemail@email.org}  

\author[orcid=0000-0003-4905-7801,sname='Gupta']{Rahul Gupta}
\affiliation{NASA Goddard Space Flight Center}
\email[hide]{noemail@email.org}  

\author[orcid=0000-0002-8585-0084,sname='Wilson-Hodge']{Colleen A. Wilson-Hodge}
\affiliation{NASA Marshall Space Flight Center}
\email[hide]{noemail@email.org}  

\author[orcid=0009-0003-3480-8251,sname='Cleveland']{William Cleveland}
\affiliation{Universities Space Research Association}
\email[hide]{noemail@email.org}  

\author[sname='Sitarski']{Breann Sitarski}
\affiliation{NASA Goddard Space Flight Center}
\email[hide]{noemail@email.org}  

\author[orcid=0000-0002-7876-7362,sname='Chand']{Vikas Chand}
\affiliation{Department of Physics \& Astronomy, Louisiana State University}
\email[hide]{noemail@email.org}  

\author[orcid=0000-0003-4768-7586,sname='Margutti']{Raffaella Margutti}
\affiliation{Department of Physics, University of California, Berkeley}
\affiliation{Department of Astronomy, University of California, Berkeley}
\email[hide]{noemail@email.org}  

\author[orcid=0000-0002-0587-7042,sname='Goldstein']{Adam Goldstein}
\affiliation{Universities Space Research Association}
\email[hide]{noemail@email.org}  

\author[orcid=0000-0002-2445-5275,sname='Foley']{Ryan J. Foley}
\affiliation{Astronomy \& Astrophysics Department, University of California, Santa Cruz}
\email[hide]{noemail@email.org}  

\author[orcid=0000-0002-4394-4138,sname='Sharma']{Vidushi Sharma}
\affiliation{Department of Physics, University of Maryland, Baltimore County}
\affiliation{NASA Goddard Space Flight Center}
\email[hide]{noemail@email.org}  

\author[orcid=0000-0001-9108-573X,sname='Yang']{Yi-Jung Yang}
\affiliation{Center for Astrophysics and Space Science, New York University Abu Dhabi}
\email[hide]{noemail@email.org}  

\author[orcid=0000-0001-9935-8106,sname='Bissaldi']{Elisabetta Bissaldi}
\affiliation{Dipartimento Interateneo di Fisica ``Michelangelo Merlin'', Politecnico di Bari}
\affiliation{INFN Bari}
\email[hide]{noemail@email.org}  

\author[orcid=0000-0002-2666-728X,sname='Miller']{M. Coleman Miller}
\affiliation{Department of Astronomy, University of Maryland}
\email[hide]{noemail@email.org}  

\author[orcid=0000-0001-9227-8349,sname='Pritchard']{T. A. Pritchard}
\affiliation{Department of Astronomy, University of Maryland}
\affiliation{NASA Goddard Space Flight Center}
\email[hide]{noemail@email.org}  

\author[orcid=0000-0003-4102-380X,sname='Sand']{D. J. Sand}
\affiliation{Department of Astronomy and Steward Observatory, University of Arizona}
\email[hide]{noemail@email.org}  

\author[orcid=0000-0002-2184-6430,sname='Ahumada']{Tomas Ahumada}
\affiliation{NSF NOIRLab}
\email[hide]{noemail@email.org}  

\author[orcid=0000-0002-4744-9898,sname='Racusin']{Judith Racusin}
\affiliation{NASA Goddard Space Flight Center}
\email[hide]{noemail@email.org}  

\author[orcid=0000-0003-4537-3575,sname='Franz']{Noah Franz}
\affiliation{Department of Astronomy and Steward Observatory, University of Arizona}
\email[hide]{noemail@email.org}  

\author[orcid=0000-0002-8297-2473,sname='Alexander']{Kate D. Alexander}
\affiliation{Department of Astronomy and Steward Observatory, University of Arizona}
\email[hide]{noemail@email.org}  

\author[orcid=0000-0001-6350-8168,sname='Mockler']{Brenna Mockler}
\affiliation{Department of Physics and Astronomy, University of California, Davis}
\email[hide]{noemail@email.org}  

\author[orcid=0000-0003-4253-656X,sname='Howell']{D. Andrew Howell}
\affiliation{Las Cumbres Observatory}
\affiliation{Department of Physics, University of California, Santa Barbara}
\email[hide]{noemail@email.org}  

\author[orcid=0000-0003-1309-2904,sname='Quintana']{Elisa Quintana}
\affiliation{NASA Goddard Space Flight Center}
\email[hide]{noemail@email.org}  

\author[orcid=0000-0001-8551-2002,sname='Hu']{Chin-Ping Hu}
\affiliation{Department and Graduate Institute of Physics, National Changhua University of Education}
\email[hide]{noemail@email.org}  

\author[orcid=0000-0002-8977-1498,sname='Andreoni']{Igor Andreoni}
\affiliation{Department of Physics and Astronomy, University of North Carolina at Chapel Hill}
\email[hide]{noemail@email.org}  

\author[orcid=0000-0001-9556-7576,sname='Hristov']{Boyan Hristov}
\affiliation{Center for Space Plasma and Aeronomic Research, University of Alabama in Huntsville}
\email[hide]{noemail@email.org}

\begin{abstract}

Time-Domain and Multi-Messenger Astrophysics (TDAMM) is entering a discovery-rich but
follow-up-limited era, creating an urgent need for responsive, multiwavelength
space-based capabilities. The Hydra constellation is a concept for a proliferated space architecture for time-domain astrophysics. The constellation would act as a disaggregated observatory composed of coordinated, relatively low-cost spacecraft that collectively provide capabilities traditionally concentrated within a single large mission. The architecture
would combine persistent wide-field gamma-ray monitoring, wide-field and focused
X-ray observations, and rapid-response ultraviolet, optical, and infrared imaging and
spectroscopy. The constellation would both discover high-energy transients and respond to
external alerts from gravitational-wave detectors, neutrino observatories, and ground-
and space-based surveys, using low-latency communications, automated event
prioritization, and community coordination frameworks to rapidly assign observing
resources. A proliferated architecture would offer operational advantages over a single
larger mission, including simultaneous observations of multiple targets, graceful
degradation following individual spacecraft failures, recurring technology refresh, and
opportunities for commercial, international, and philanthropic contributed nodes to join the
network. The constellation would address fundamental questions concerning cosmic
accelerators, the origin and evolution of the elements, the behavior of matter at extreme
density, and the nature of dark energy through gravitational-wave standard sirens. This white paper presents the Hydra concept description that was submitted to NASA's ASTRA initiative for consideration by the Cosmic Origins Program Analysis Group (CoPAG) and Physics of the Cosmos Program Analysis Group (PhysPAG).

\end{abstract}


\section{Science Investigation} \label{sec:science}

Time-Domain and Multi-Messenger Astrophysics (TDAMM) is entering a fundamentally new phase characterized by an unprecedented increase in the rate and diversity of astrophysical transient detections. The field is transitioning from a discovery-limited to a follow-up-limited era, driven by major investments across electromagnetic, gravitational-wave, and neutrino observatories. New and upcoming facilities such as the Vera C. Rubin Observatory, the Nancy Grace Roman Space Telescope, the Argus Array, and other wide-field surveys will produce a deluge of time-domain alerts, reaching millions of events per night. Simultaneously, next generation gravitational-wave networks and neutrino observatories will significantly increase the detection rates of non-electromagnetic messengers. This discovery-rich landscape creates a major scientific opportunity, but realizing its full potential will require responsive, multiwavelength, space-based facilities that can identify and follow up events quickly, flexibly, and coherently.

The Hydra constellation is a proliferated space architecture for time-domain astrophysics that addresses this need through a coordinated suite of relatively low-cost space telescopes that together exceed the scientific capability of an integrated TDAMM probe-class observatory. Rather than implementing a single monolithic observatory, the constellation disaggregates the major functional elements of a rapid transient mission across multiple spacecraft: all-sky gamma-ray monitoring for prompt discovery of high-energy transients, multiple rapid-response X-ray telescopes for arcsecond-scale localization and characterization, and multiple rapid-response ultraviolet, optical, and infrared (UVOIR) telescopes for early counterpart identification and time-series photometric, spectroscopic, and polarimetric observations. 

The core science gap addressed by Hydra is the impending loss or degradation of high-energy monitoring and rapid-response capabilities combined with the explosive growth in space and ground-based alert streams. The Swift observatory demonstrated the transformative value of coupling wide-field gamma-ray discovery to rapid X-ray, UV, and optical follow-up. Fermi-GBM has provided all-sky gamma-ray monitoring and has been central to gravitational-wave counterpart searches. However, the current high-energy fleet is aging, Swift’s long-term future is uncertain, and the community faces a potential gap in all-sky gamma-ray discovery, prompt X-ray observations, and space-based, early-time UVOIR follow-up just as Rubin, Argus, and other surveys will produce a rich stream of optical transients, with high-impact sources requiring immediate multiwavelength response.

Hydra would close this gap by providing a persistent, low-latency, multiwavelength response system optimized for transient events. It would discover and localize short and long gamma-ray bursts, fast high-energy transients, magnetar giant flares, tidal disruption events with prompt high-energy emission, relativistic supernova shock breakouts, and other explosive events. It would also respond to external triggers from gravitational-wave detectors, neutrino observatories, as well as discoveries made by Rubin, Roman, Argus, and radio facilities. The mission would serve as both a discovery and follow-up engine by generating its own high-energy triggers and rapidly tasking its X-ray and UVOIR elements, while also accepting high-priority external alerts and target of opportunity (ToO) requests for coordinated follow-up. 

The primary science objectives (SO) that would be addressed by Hydra are:
\newline
\newline
\textbf{SO1: How do cosmic accelerators work and what are they accelerating?} The universe produces explosions and outbursts capable of accelerating matter and radiation to extreme energies, but the engines that power many of these events remain poorly understood [1][2][3]. High-energy, time-domain observations can reveal how massive stars die, jets are launched or fail, shocks emerge from stellar envelopes, and how relativistic outflows produce gamma-ray bursts, X-ray flashes, fast X-ray transients, and orphan afterglows [4]. Sensitive wide-field X-ray and gamma-ray transient discovery combined with rapid X-ray and UVOIR follow-up will connect these phenomena to their progenitors, environments, and energy sources. This will allow us to determine whether observationally distinct classes like low-luminosity GRBs, ultra-long GRBs, luminous fast-blue optical transients, shock breakout events, tidal disruption events, magnetar flares and their link to fast radio bursts, compact-object mergers, and high-energy neutrino sources, represent separate physical channels or different observational views of a smaller set of underlying engines [5][6][7][8]. By surveying the dynamic high-energy sky across cosmic time, including high-redshift GRBs, Hydra would identify the sources of cosmic acceleration, determine what particles and outflows they produce, and reveal how compact objects, massive stars, and black holes convert gravitational, magnetic, rotational, and accretion energy into relativistic phenomena [2][9].
\newline
\newline
\textbf{SO2: What is the elemental enrichment history of the universe?} The chemical history of the universe has been written in part by explosive transients. High-redshift gamma-ray bursts offer luminous backlights for probing the first generations of massive stars and the earliest production of metals when the universe was still young [10]. At later times, the heaviest elements, from iron-peak nuclei to the r-process elements extending toward uranium, are forged in rare, extreme events such as neutron-star mergers [11] and extreme magnetar explosions [12]. Discovery with sensitive high-energy monitors, followed by rapid UVOIR characterization of kilonovae and related thermal transients, will measure elemental yields, identify the dominant production channels, and map their production rates across cosmic time [13]. These observations will determine how the universe became chemically enriched and when the first planets with environments capable of supporting life as we know it were formed [14].
\newline
\newline
\textbf{SO3: What are the new states of matter at high density and temperature?} Astronomy provides one of the only ways to probe matter at densities beyond the reach of terrestrial laboratories. A central goal is to determine the neutron-star equation of state, a foundational input to nuclear physics and our understanding of the properties of the heaviest nuclei [15][16]. Multimessenger observations of neutron-star mergers, discovered with gravitational-wave detectors and high-energy monitors and subsequently characterized across the electromagnetic spectrum, can constrain the maximum mass of a neutron star, a uniquely powerful probe of the densest stable matter in the universe [17][18][19][20]. Precise X-ray observations of Galactic neutron stars can measure masses and radii, providing complementary constraints on the pressure-density relation of ultra-dense matter [21][22][23]. Together, these observations may reveal the densities at which hadronic matter transitions to deconfined quark matter, mapping a phase transition that remains inaccessible to current terrestrial experiments [24][25].
\newline
\newline
\textbf{SO4: What is the nature of dark energy?} Is dark energy a cosmological constant, or does it evolve with cosmic time? If it is dynamical, what physical mechanism drives its behavior [26][27]? Roman, Rubin, and Euclid are designed to address this question through complementary measurements of cosmic expansion and structure growth [28][29]. With the next generation of GW observatories, GW-detected “standard sirens” will offer an independent path with potentially minimal systematic uncertainties: General Relativity provides absolute distance measurements from gravitational-wave observations of neutron-star mergers, while electromagnetic counterparts provide localizations, host-galaxy identifications, and redshifts [30][31]. Extending this method deep into the universe requires sensitive all-sky monitoring of the gamma-ray sky to discover merger counterparts, together with rapid-response X-ray, optical, and infrared telescopes to capture fading afterglows and kilonovae, identify host galaxies, and measure the redshifts needed to map cosmic expansion [32][8]. In the absence of a theoretical breakthrough on dark energy, precise measurement of the cosmic expansion will provide our best guidance as to its nature [27].


\begin{table*}[t]
    \centering
    \caption{Science objectives and the physical parameters and observables that are needed to address them.}
    \label{Table:brokers}

    \small
    \setlength{\tabcolsep}{4pt}
    \renewcommand{\arraystretch}{1.1}

    \begin{tabular}{
        |>{\raggedright\arraybackslash}p{0.23\textwidth}
        |>{\raggedright\arraybackslash}p{0.23\textwidth}
        |>{\raggedright\arraybackslash}p{0.23\textwidth}
        |>{\raggedright\arraybackslash}p{0.23\textwidth}|
    }
        \hline

        \textbf{Science Objectives} &
        \textbf{Physical Parameters} &
        \textbf{Observables} &
        \textbf{Potential Challenges} \\
        \hline

        \textbf{SO1: Determine how cosmic accelerators work and what they are accelerating.} &
        Jet energy and jet structure, Lorentz factor, ejecta mass and velocity,
        shock breakout radius and temperature, magnetic-field strength, accretion
        rate, compact-object mass and spin, particle acceleration efficiency,
        circumburst density, redshift distribution. &
        X-ray and gamma-ray transient detections; prompt light curves, spectra,
        fluence, duration, hardness, and localization; early X-ray and UVOIR
        afterglow light curves; host-galaxy properties and redshifts; gravitational
        wave and neutrino alert associations. &
        Rare, fast-fading, diverse events require high-duty-cycle wide-field
        discovery, rapid localization, and coordinated follow-up. Interpretation
        of individual events is limited by viewing-angle, progenitor, and ambiguity
        between classes of engine. \\
        \hline

        \textbf{SO2: Determine the elemental enrichment history of the universe.} &
        Redshift, metallicity, elemental abundance patterns, ejecta mass, ejecta
        velocity, opacity, kilonova temperature and luminosity, nucleosynthetic
        yield, merger and explosion rates, host-galaxy age and environment. &
        High-redshift GRB afterglow spectra probing early galaxies; kilonova UVOIR
        light curves, colors, and spectra; host redshifts \& metallicities;
        neutron-star merger, magnetar-powered explosion, and related transient
        rates over cosmic time. &
        Highest redshift events are faint and rare, and kilonovae evolve rapidly
        and are intrinsically dim. Elemental yield inference depends on radiative
        transfer modeling, opacity uncertainties, and the ability to obtain rapid
        IR follow-up and spectroscopy before the transient fades. \\
        \hline

        \textbf{SO3: Determine the new states of matter at high density and temperature.} &
        Neutron-star mass, radius, compactness, maximum stable mass, post-merger
        remnant lifetime, pressure density relation, temperature, spin, magnetic
        field, nuclear symmetry energy, TDEs, magnetar energy injection, extinction. &
        Prompt gamma-ray counterparts or lack thereof; X-ray afterglows and plateaus;
        kilonova properties tied to remnant lifetime and ejecta mass; mass-radius
        constraints from pulse profiles, thermonuclear bursts, and accretion-powered
        systems. &
        Dense-matter observables are indirect and model dependent. Constraints
        require joint gravitational wave, gamma-ray, X-ray, and UVOIR interpretation,
        while short-lived post-merger emission and neutron-star atmosphere,
        magnetic-field, and geometry systematics can limit mass-radius precision. \\
        \hline

        \textbf{SO4: Determine the nature of dark energy using standard sirens.} &
        Luminosity distance, redshift, Hubble constant, dark-energy equation-of-state
        parameters, merger rate evolution, inclination angle, host-galaxy association
        probability. &
        Gravitational-wave distances from NS mergers; gamma-ray counterparts;
        afterglows and kilonova light curves; arcsecond localizations; host redshifts;
        standard siren statistics across cosmic time. &
        Standard sirens require gravitational wave detections, electromagnetic
        counterparts, and secure host redshifts. Large localizations, rapid fading,
        viewing-angle degeneracies. \\
        \hline

    \end{tabular}
\end{table*}

\section{Instrumentation Description} \label{sec:instrumentation}

Hydra is conceived as a proliferated space architecture rather than a single monolithic observatory. The mission would consist of coordinated spacecraft layers, with each layer employing mature, flight-proven instrument concepts wherever possible to reduce development timescales. The baseline architecture includes gamma-ray discovery, X-ray discovery and follow-up, and UVOIR follow-up elements operating as an integrated time-domain observatory.

The gamma-ray discovery layer would provide continuous, wide-field monitoring for high-energy transients using modular scintillation detectors with heritage from Fermi-GBM and StarBurst [33]. It could be implemented as hosted or free-flying nodes in multiple orbits, or as a multi-detector payload near the Sun–Earth L2 or L4/L5 points, where it would benefit from unocculted sky coverage and long IPN-like triangulation baselines. A recently commissioned NASA study to place hosted gamma-ray detectors on the future Habitable Worlds Observatory provides one possible path to such an L2 implementation.  Across these options, the gamma-ray nodes would issue low-latency alerts containing trigger time, localization, and preliminary classification to the rest of the constellation.

The X-ray element would provide both wide-field discovery and localization, as well as  focused follow-up. The wide-field component would search for X-ray transients that may be missed or only coarsely localized by gamma-ray instruments. Notional implementations could include soft X-ray compact micropore-optics telescopes [34], coded-aperture imagers such as BlackCAT [35] or the Wide Field Monitor proposed for the Strobe-X probe [36] concept. The focused follow-up component would provide deeper, higher-angular-resolution observations, with emphasis on rapid slews, arcsecond-scale localization, early afterglow properties, light-curve evolution, and basic spectral diagnostics. This component could use focusing X-ray telescopes with CCD, CMOS, or hybrid detectors, including narrow-field telescopes drawing on NICER [37] or IXPE-like [38] focusing optics. A representative bandpass would be approximately 0.3–10 keV, with final sensitivity and field-of-view requirements set by science-yield trades.

The UVOIR element provides early photometric characterization, counterpart identification, color evolution and spectroscopic measurements. A notional implementation would use multiple moderate-aperture telescopes with rapid pointing capabilities. The UV capability is especially valuable for early shock breakout, supernova cooling emission, tidal disruption events, and young kilonova emission, while optical and near-infrared bands are critical for kilonova evolution, high-redshift GRBs, and classification of Rubin-discovered transients. A possible implementation path is a fleet of Pandora-like observatories [39], or similar commercially available telescopes. Pandora demonstrated that a half-meter-class space telescope can be packaged in a compact form factor and can support simultaneous visible photometry and near-infrared spectroscopy using a dual-channel payload. In this implementation, one tranche could deploy visible/NIR nodes, while complementary UV-sensitive nodes could be added in the same or later tranches to provide the full UVOIR capability. Key trades to study would include aperture size, field of view, wavelength coverage, and the relative value of wide-field imaging, slit spectroscopy, and integral-field spectroscopy. The proliferated UVOIR approach enables more frequent target access, simultaneous monitoring of multiple transients, and flexible observing strategies that are difficult for a small number of highly oversubscribed facilities to provide. 

\section{Mission Implementation} \label{sec:mission}

The Hydra constellation takes inspiration from the Proliferated Warfighter Space Architecture [40] currently being pursued by the Space Development Agency (SDA), where traditional reliance on a small number of exquisite, high-cost satellites is being replaced by a resilient, low-Earth-orbit architecture composed of many smaller, more rapidly developed spacecraft. This same architectural philosophy offers a compelling path toward a next-generation TDAMM capability: rather than building a single monolithic replacement for missions such as Swift or Fermi, Hydra would distribute discovery, localization, and follow-up capabilities across a larger network of modular spacecraft. This approach would provide graceful degradation if individual nodes fail, increased sky coverage and observing responsiveness, enable multiple high-priority transients to be followed simultaneously, and allow the architecture to be replenished or upgraded over time. As new technologies and commercial spacecraft capabilities mature, later nodes could incorporate these advances without requiring the redesign of an entire flagship-class observatory. 

Hydra would support self-triggered, externally triggered, and survey mode observing. In the self-triggered mode, the gamma-ray monitor detects and characterizes a high-energy transient, issues a low-latency alert, and initiates rapid X-ray observations to refine the localization, measure early afterglow behavior, and identify or constrain UVOIR counterparts, with rapid data release to enable broader community response. In the externally triggered mode, Hydra ingests alerts from discovery engines like gravitational-wave detectors, neutrino observatories, wide-field optical and radio surveys, and ToO requests. An automated AI/ML system, in conjunction with new coordination frameworks such as ACROSS [41], would triage each event using trigger properties, observability, and resource availability, and route high-priority targets to the X-ray and UVOIR nodes with human-in-the-loop approval for disruptive or ambiguous cases. Wide-area surveys would utilize non-allocated time.

The constellation’s communications architecture would be designed around low-latency alert delivery, rapid command dissemination, and prompt data return. Options include new commercial relay services, optical communications, and inter-satellite crosslinks. Hydra would leverage commercial and partner capabilities by using the model established by the PWSA program, which uses recurring tranches, multiple vendors, and published interoperability standards to allow independently procured space and ground elements to operate as part of a common architecture. In particular, the SDA optical communications [42] and networking protocols illustrate how published standards can reduce vendor lock-in, support competitive procurement, and allow for commercial, international, or philanthropic-contributed nodes to join the architecture over time. A traditional mission lifetime requirement may not fully capture the Hydra concept. Individual nodes might be designed for three to five years, while the architecture itself would be designed for ten or more years of sustained capability through replenishment.

The ASTRA incubator should study the Hydra constellation as a valuable new mission concept: a strategic distributed observatory. Key study products would include: (1) Science trades and capability assessments for gamma-ray, X-ray, and UVOIR instrument implementations; (2) Constellation architecture trades as to number of nodes, orbit, communications, and response latency; (3) Technology-readiness assessments identifying which elements can be implemented with existing flight heritage and which require focused technology maturation.

\section{References} \label{sec:refs}

\noindent[1] Woosley \& Bloom 2006, Annual Review of Astronomy and Astrophysics, 44, 507. \newline
[2] Kumar \& Zhang 2015, Physics Reports, 561, 1. \newline
[3] Mészáros et al. 2019, Nature Reviews Physics, 1, 585. \newline
[4] Nakar \& Sari 2012, The Astrophysical Journal, 747, 88. \newline
[5] Levan et al. 2014, The Astrophysical Journal, 781, 13. \newline
[6] Margutti et al. 2019, The Astrophysical Journal, 872, 18. \newline
[7] De Colle \& Lu 2020, New Astronomy Reviews, 89, 101538. \newline
[8] Abbott et al. 2017, The Astrophysical Journal Letters, 848, L12. \newline
[9] Fryer et al. 2022, The Astrophysical Journal, 929, 111. \newline
[10] Wang et al. 2012, The Astrophysical Journal, 760, 27. \newline
[11] Metzger et al. 2010, Monthly Notices of the Royal Astronomical Society, 406, 2650. \newline
[12] Patel et al. 2025, The Astrophysical Journal Letters, 984, L29. \newline
[13] Kasen et al. 2017, Nature, 551, 80. \newline
[14] Johnson \& Li 2012, The Astrophysical Journal, 751, 81. \newline
[15] Lattimer \& Prakash 2007, Physics Reports, 442, 109. \newline
[16] Lattimer 2021, Annual Review of Nuclear and Particle Science, 71, 433. \newline
[17] Margalit \& Metzger 2017, The Astrophysical Journal Letters, 850, L19. \newline
[18] Rezzolla et al. 2018, The Astrophysical Journal Letters, 852, L25. \newline
[19] Radice et al. 2018, The Astrophysical Journal Letters, 852, L29. \newline
[20] Bauswein 2021, arXiv e-prints, arXiv:2103.16371. \newline
[21] Watts et al. 2016, Reviews of Modern Physics, 88, 021001. \newline
[22] Miller et al. 2019, The Astrophysical Journal Letters, 887, L24. \newline
[23] Raaijmakers et al. 2021, The Astrophysical Journal Letters, 918, L29. \newline
[24] Annala et al. 2020, Nature Physics, 16, 907. \newline
[25] Annala et al. 2023, Nature Communications, 14, 8451. \newline
[26] Peebles \& Ratra 2003, Reviews of Modern Physics, 75, 559. \newline
[27] Frieman et al. 2008, Annual Review of Astronomy and Astrophysics, 46, 385. \newline
[28] Spergel et al. 2015, arXiv e-prints, arXiv:1503.03757. \newline
[29] Mellier et al. 2025, Astronomy \& Astrophysics, 697, A1. \newline
[30] Schutz 1986, Nature, 323, 310. \newline 
[31] Chen et al. 2018, Nature, 562, 545. \newline
[32] Goldstein et al. 2017, The Astrophysical Journal Letters, 848, L14. \newline
[33] Woolf et al. 2024, Nuclear Instruments and Methods in Physics Research Section A, 1064. \newline
[34] Yuan et al. 2018, Proceedings of SPIE, 10699, 1069925. \newline
[35] Chattopadhyay et al. 2018, Proceedings of SPIE, 10699, 106995S. \newline
[36] Ray et al. 2019, Bulletin of the American Astronomical Society, 51, 231. \newline
[37] Gendreau et al. 2016, Proceedings of SPIE, 9905, 99051H. \newline
[38] Soffitta et al. 2021, The Astronomical Journal, 162, 208. \newline
[39] Barclay et al. 2025, IEEE Aerospace Conference, 1, 14. \newline
[40] \url{https://www.sda.mil/wp-content/uploads/2024/05/Transport-Layer_distro-A_FINAL.pdf} \newline
[41] Humensky et al. 2024, Frontiers in Astronomy and Space Sciences, 11, 1401785. \newline
[42] Space Development Agency 2025, USSF Standard, 9100-001-11, v3.2.0. \newline


\end{document}